\documentclass[sigplan,screen,nonacm]{acmart}
\usepackage{multirow}
\usepackage[htt]{hyphenat}

\usepackage{listings}
\definecolor{codegreen}{rgb}{0,0.6,0}
\definecolor{codegray}{rgb}{0.5,0.5,0.5}
\definecolor{codepurple}{rgb}{0.58,0,0.82}
\definecolor{backcolour}{rgb}{0.95,0.95,0.92}

\lstdefinestyle{mystyle}{
    backgroundcolor=\color{backcolour},   
    commentstyle=\color{codegreen},
    keywordstyle=\color{magenta},
    numberstyle=\tiny\color{codegray},
    stringstyle=\color{codepurple},
    basicstyle=\ttfamily\footnotesize,
    breakatwhitespace=false,         
    breaklines=true,                 
    captionpos=b,                    
    keepspaces=true,                 
    numbers=left,                    
    numbersep=5pt,                  
    showspaces=false,                
    showstringspaces=false,
    showtabs=false,                  
    tabsize=2
}

\AtBeginDocument{%
  }

\setcopyright{none}

\begin{document}

\title[Rust for Secure Backend Development]{Rust for Secure Backend Development: A Critical Review and Extended Vulnerability Comparison with Node.js and Django}


\author{Md Zarzees Uddin Shah Chowdhury}
\affiliation{%
  \institution{Virginia Tech}
  \city{Blacksburg, VA}
  \country{USA}}
\email{zarzees5075@vt.edu}

\author{Rabib Jahin Ibn Momin}
\affiliation{%
  \institution{Bangladesh University of Engineering and Technology}
  \city{Dhaka}
  \country{Bangladesh}}

\author{Rifat Shahriyar}
\affiliation{%
  \institution{Bangladesh University of Engineering and Technology}
  \department{Department of Computer Science and Engineering}
  \city{Dhaka}
  \country{Bangladesh}}

\renewcommand{\shortauthors}{Chowdhury, Momin, and Shahriyar}

\begin{abstract}
The Rust programming language is widely credited with eliminating entire classes of memory-safety and concurrency vulnerabilities, but the security implications of adopting it in practice extend well beyond memory safety. This paper presents a critical review of prior work on Rust's security posture in industrial settings~\cite{gasiba2023rust}, and extends that analysis in a direction the original study did not cover: backend web development. We first assess the strengths and limitations of the existing vulnerability classification of Rust against C, C++, and Java under the SANS Top 25, OWASP Top 10, and the 19 Deadly Sins of Software Security frameworks, identifying gaps including limited empirical validation, a small interview sample, and the absence of a secure development lifecycle discussion. We then contribute an original comparison of Rust against Node.js and Django using the same three-level classification (Rare and Difficult, Safeguarded, Unprotected), supported by side-by-side code experiments for out-of-bounds writes (CWE-787), use-after-free (CWE-416), and race conditions (CWE-362). Our results indicate that Rust's compile-time guarantees dominate at the systems layer, while managed backend frameworks offer stronger built-in defenses at the application layer, suggesting that Rust adoption in web contexts requires complementary safeguards rather than reliance on language-level safety alone.
\end{abstract}



\maketitle

\section{Introduction}
The Rust programming language, emerging in 2010, has rapidly gained prominence in industrial software development due to its emphasis on \emph{memory safety} and \emph{concurrency}. With adoption in critical systems such as the Linux Kernel and Android, Rust positions itself as a robust alternative to traditional languages like C and C++. Its compile-time guarantees against memory corruption, data races, and null pointer dereferences address long-standing vulnerabilities pervasive in systems programming. However, while Rust’s safety mechanisms are widely celebrated, a systematic evaluation of its security posture in industrial contexts—particularly in critical infrastructure adhering to standards like \emph{IEC 62443} remains underexplored.  

Gasiba and Amburi~\cite{gasiba2023rust} investigate Rust’s efficacy in mitigating software vulnerabilities compared to C, C++, and Java, languages entrenched in industrial practice. Despite Rust’s advancements, conflating \emph{safety} with \emph{security} risks overlooking non-memory-related flaws, such as logic errors, insecure configurations, and web-layer vulnerabilities (e.g., Cross-Site Scripting). Their research addresses this gap through a multi-faceted methodology:
\begin{itemize}
    \item \textbf{Literature Review}: Analyzing academic and gray literature (2010--2023) on Rust’s security model, vulnerabilities, and tooling.  
    \item \textbf{Expert Interviews}: Gathering insights from industry professionals with over a decade of experience in secure software development.  
    \item \textbf{Vulnerability Mapping}: Aligning Rust’s protections with frameworks like SANS Top 25, OWASP Top 10, and the \emph{19 Deadly Sins of Software Security}.  
    \item \textbf{Tool Analysis}: Evaluating Static Application Security Testing (SAST) tools for Rust against mature ecosystems like Java and C++.  
\end{itemize}

Their findings reveal that while Rust eliminates 24\% of SANS Top 25 vulnerabilities (e.g., buffer overflows, race conditions) through language design, 48\% remain unaddressed (e.g., path traversal, insecure authentication). Notably, vulnerabilities often stem from misuse of \texttt{unsafe} code blocks or design-level oversights unrelated to memory safety. Compared to C/C++ (no inherent protections) and Java (garbage-collected memory safety), Rust strikes a balance between low-level control and safety but necessitates complementary secure coding practices.  

In this paper we critically review that study and extend it. Our contributions are twofold. First, we assess the strengths and limitations of their methodology and vulnerability classification (Section~\ref{sec:critical}), and reproduce representative code experiments to verify the classification decisions they report. Second, we extend the comparison beyond systems languages to backend web development, evaluating Rust against Node.js and Django under the same three-level scheme (Section~\ref{sec:experiment}). As industries increasingly prioritize cybersecurity resilience, understanding Rust’s role—as both a safeguard and a potential risk—is critical for shaping future secure development paradigms.

\section{Related Work}
Research on Rust’s security has evolved alongside its industrial adoption. Sible et al.~\cite{Sible2022} conducted a foundational analysis of Rust’s memory and concurrency safety, highlighting limitations such as memory leaks and the need for holistic security practices. Wassermann et al.~\cite{Wassermann2023} expanded this by emphasizing vulnerabilities arising from misaligned design assumptions and advocating for ecosystem maturity. Their work stresses the importance of analyzing real-world vulnerabilities, even without source code access, to mitigate risks in Rust-based systems.  

Qin et al.~\cite{Qin2020} empirically studied unsafe code usage in Rust projects, revealing that while developers minimize unsafe blocks, memory-safety bugs often stem from interactions between safe and unsafe code. Their findings underscore the complexity of Rust’s lifetime system and the risks of integrating unsafe code for performance. Xu et al.~\cite{Xu2021} further analyzed all Rust common vulnerability and exposures (CVEs), concluding that memory-safety challenges persist despite Rust’s compile-time guarantees, particularly when developers bypass safety mechanisms.  

\paragraph{Security Standards and Industrial Guidelines.}  
Industrial adoption of Rust necessitates alignment with standards like \emph{IEC 62443}~\cite{IEC62443} for critical infrastructure and \emph{ISO/IEC TR 24772}~\cite{ISO24772} for secure coding. The \emph{Common Weakness Enumeration (CWE)}~\cite{CWE} framework provides a unified taxonomy of software weaknesses, while ANSSI’s guidelines for Rust~\cite{ANSSI2023} offer tailored rules to address its safety-security interplay.  

\paragraph{Tools and Ecosystem Maturity.}  
The Rust ecosystem includes tools like \texttt{RustSec}~\cite{RustSec2023} for tracking crate vulnerabilities and SAST tools listed in \texttt{Analysis Tools}~\cite{AnalysisTools2023}. Google’s integration of Rust into Android~\cite{Google2022} exemplifies its industrial use to mitigate memory-safety bugs. However, maturity gaps persist compared to Java/C++ ecosystems, which leverage tools like \texttt{SonarQube}~\cite{SonarQube2023} for compliance.  

\paragraph{Secure Coding Practices and Awareness.}  
Gasiba et al.~\cite{Gasiba2021} identified a gap between developers’ intent to follow secure coding guidelines and their practical knowledge. Their work inspired platforms like \emph{Sifu}~\cite{Gasiba2020}, which gamifies cybersecurity education. Bagnara et al.~\cite{Bagnara2022} explored the undecidability of secure coding rules (e.g., input validation), emphasizing that no language can fully automate vulnerability prevention. Jacoby et al.~\cite{Jacoby1971} validated the three-point vulnerability classification (RD/SG/UP) by demonstrating the sufficiency of Likert-style scales for such analyses.  

\paragraph{Rust CVEs and Mitigations.}  
The Rust Security Response WG addresses vulnerabilities like \texttt{CVE-2021-42574}~\cite{RustCVE2021} (Unicode trojans) and \texttt{CVE-2022-46176}~\cite{RustCVE2023} (Cargo SSH vulnerabilities). Amburi’s PoC code~\cite{Amburi2023} demonstrates exploitable weaknesses (e.g., SQL injection) even in Rust, reinforcing the need for complementary safeguards.  

Collectively, these works contextualize Rust’s strengths but leave gaps in systematic vulnerability analysis across industrial standards. The study we review~\cite{gasiba2023rust} bridges this by mapping Rust’s protections to SANS Top 25, OWASP Top 10, and the \emph{19 Deadly Sins}~\cite{Howard2005}, while addressing tooling and awareness challenges specific to industrial adoption. None of this prior work, however, evaluates Rust against the managed backend stacks that dominate web development, which is the gap our extension in Section~\ref{sec:experiment} addresses.

\section{Methodology of the Reviewed Study}
\label{sec:reviewed-method}
We summarize the methodology of Gasiba and Amburi~\cite{gasiba2023rust} here, since our critical analysis in Section~\ref{sec:critical} assesses it directly and our extension in Section~\ref{sec:experiment} reuses its classification scheme. Their study evaluates Rust’s security capabilities through a multi-phase approach, combining qualitative and quantitative analyses, and addresses two research questions:
\textbf{RQ1:} \emph{To what extent does Rust mitigate vulnerabilities compared to C, C++, and Java?}  
\textbf{RQ2:} \emph{How effective are existing tools and practices for secure Rust development in industrial contexts?}  

\subsection{Research Design}
The methodology comprises four stages:  

\begin{itemize}
    \item \textbf{Literature Review}: They analyzed academic and gray literature (2010--2023) from ACM, IEEE Xplore, and Google Scholar using keywords like \texttt{"Rust Security"} and \texttt{"SAST Tools"}. This included foundational works by Qin et al.~\cite{Qin2020} on unsafe code and Gasiba et al.~\cite{Gasiba2021} on secure coding awareness.  
    
    \item \textbf{Expert Interviews}: Semi-structured interviews were conducted with five industry security experts (10+ years of experience) and two open-source Rust contributors. Discussions focused on Rust’s adoption challenges, vulnerability patterns, and tooling gaps.  
    
    \item \textbf{Vulnerability Mapping}: They mapped Rust’s protections to three frameworks:  
    \begin{enumerate}
        \item \emph{SANS Top 25}~\cite{SANSTop25} and \emph{CWE}~\cite{CWE} for system-level weaknesses.
        \item \emph{OWASP Top 10}~\cite{OWASP2021} for application-layer vulnerabilities.
        \item \emph{19 Deadly Sins of Software Security}~\cite{Howard2005} for design flaws.
    \end{enumerate}
    
    \item \textbf{Static Analysis}: They evaluated Rust’s SAST tools (e.g., \texttt{Clippy}, \texttt{RustSec}) against Java (\texttt{SonarQube}~\cite{SonarQube2023}) and C++ (\texttt{cppcheck}) ecosystems.  
\end{itemize}

\subsection{Vulnerability Classification}
They categorized vulnerabilities using a three-point scale inspired by Jacoby et al.~\cite{Jacoby1971}:  
\begin{enumerate}
    \item \textbf{Rare and Difficult (RD)}: Fully mitigated by Rust’s design (e.g., buffer overflows~\cite{Xu2021}) unless \texttt{unsafe} blocks are used.  
    \item \textbf{Safeguarded (SG)}: Partially addressed by Rust or dependent on libraries (e.g., SQL injection via \texttt{rusqlite}~\cite{Amburi2023}).  
    \item \textbf{Unprotected (UP)}: Require external mitigations (e.g., insecure authentication~\cite{ANSSI2023}).  
\end{enumerate}

\subsection{Validation}
The authors developed Proof-of-Concept (PoC) code~\cite{Amburi2023} demonstrating vulnerabilities in Rust (e.g., command injection, TOCTOU). These were analyzed using SAST tools to assess detection efficacy. Industrial experts validated findings against IEC 62443~\cite{IEC62443} compliance requirements. 

\section{Findings of the Reviewed Study}
\label{sec:reviewed-results}
This section summarizes the results reported by Gasiba and Amburi~\cite{gasiba2023rust}. Tables~\ref{tab:sans}, \ref{tab:comparison}, \ref{tab:owasp}, and~\ref{tab:sins} reproduce their vulnerability mappings across industrial standards, and all percentages quoted below are theirs. The code listings in Section~\ref{sec:verify-code} are our own, written to independently check two of the classification decisions they report. Our original results are presented separately in Section~\ref{sec:experiment}.

\subsection{SANS Top 25 Vulnerabilities in Rust}
Table~\ref{tab:sans} reproduces their categorization of Rust’s protection levels for the SANS Top 25 CWEs. They report that Rust fully mitigates 24\% of these vulnerabilities (e.g., buffer overflows~\cite{Xu2021}) through compile-time checks, while 48\% remain unprotected (e.g., path traversal).

\begin{table}[ht]
\centering
\caption{SANS Top 25 CWE vs. Protection Levels in Rust. Reproduced from Gasiba and Amburi~\cite{gasiba2023rust}.}
\label{tab:sans}
\begin{tabular}{|l|l|c|c|c|}
\hline
\textbf{CWE ID} & \textbf{Short Description} & \textbf{RD} & \textbf{SG} & \textbf{UP} \\ \hline
CWE-787 & Out-of-bounds Write & $\bullet$ &  & \\ \hline
CWE-79 & Cross-site Scripting &  &  & $\bullet$ \\ \hline
CWE-89 & SQL Injection &  & $\bullet$ & \\ \hline
CWE-20 & Improper Input Validation &  & $\bullet$ &  \\ \hline
CWE-125 & Out-of-bounds Read & $\bullet$ &  &  \\ \hline
CWE-78 & OS Command Injection &  &$\bullet$  &  \\ \hline
CWE-416 & Use After Free & $\bullet$ &  &  \\ \hline
CWE-22 & Path Traversal &  &  & $\bullet$ \\ \hline
CWE-352 & CSRF &  &  & $\bullet$ \\ \hline
CWE-476 & NULL Pointer Dereference &$\bullet$  &  &  \\ \hline
CWE-190 & Integer Overflow &  & $\bullet$ &  \\ \hline
CWE-798 & Improper Authentication &  & & $\bullet$   \\ \hline
CWE-259 & Use of Hard-coded Credentials &  &  & $\bullet$ \\ \hline
CWE-362 & Missing Authorization &  &  &$\bullet$  \\ \hline
CWE-77 & Command Injection &  & $\bullet$   & \\ \hline
CWE-119 & Buffer Overflow & $\bullet$  & &  \\ \hline
CWE-276 & Incorrect Default Permissions &  &  & $\bullet$  \\ \hline
CWE-918 & Server-Side Request Forgery &  &  & $\bullet$ \\ \hline
CWE-362 & Race Condition &$\bullet$  &  &  \\ \hline

CWE-611 & Improper Restriction of XXE &  &  & $\bullet$ \\ \hline
CWE-94 & Code Injection &  & $\bullet$ &  \\ \hline
\end{tabular}
\smallskip
\centering
\textit{Protection Levels:} Rare and Difficult (RD) 24\%, Safeguarded (SG) 28\%, Unprotected (UP) 48\%.
\end{table}

\subsection{Comparison with C, C++, and Java}
Table~\ref{tab:comparison} reproduces their classification for C, C++, and Java, highlighting Rust’s advantages over C/C++ (no inherent protections) and Java (partial safeguards via garbage collection). In their scheme, Rust mitigates all memory-safety CWEs (e.g., CWE-119) unless \texttt{unsafe} is used.

\begin{table}[ht]
\centering
\caption{SANS Top 25 Protection Levels in C, C++, and Java. Reproduced from Gasiba and Amburi~\cite{gasiba2023rust}.}
\label{tab:comparison}
\begin{tabular}{|l|c|c|c|c|c|c|c|c|c|}
\hline
\multirow{2}{*}{\textbf{CWE}} & \multicolumn{3}{c|}{\textbf{C}} & \multicolumn{3}{c|}{\textbf{C++}} & \multicolumn{3}{c|}{\textbf{Java}} \\ \cline{2-10}
 & RD & SG & UP & RD & SG & UP & RD & SG & UP \\ \hline
CWE-787 & &  & $\bullet$ &  & $\bullet$ &  & $\bullet$ &  &  \\ \hline
CWE-79 &  &  & $\bullet$ &  &  & $\bullet$ &  &  & $\bullet$ \\ \hline
CWE-89 &  &  & $\bullet$ &  & $\bullet$ &  &  & $\bullet$ &  \\ \hline
CWE-20 &  &  & $\bullet$ &  &  & $\bullet$ &  & $\bullet$ &  \\ \hline
CWE-125 &  &  & $\bullet$ &  &  & $\bullet$ &  & $\bullet$ &  \\ \hline
CWE-78 &  &  & $\bullet$ &  &  & $\bullet$ &  &  & $\bullet$ \\ \hline
CWE-416 &  &  & $\bullet$ &  & $\bullet$ &  & $\bullet$ &  &  \\ \hline
CWE-352 &  &  & $\bullet$ &  &  & $\bullet$ &  &  & $\bullet$ \\ \hline
CWE-434 &  &  & $\bullet$ &  &  & $\bullet$ &  &  & $\bullet$ \\ \hline
CWE-476 &  &  & $\bullet$ &  & $\bullet$ &  &  & $\bullet$ &  \\ \hline
CWE-502 &  &  & $\bullet$ &  &  & $\bullet$ &  &  & $\bullet$ \\ \hline
CWE-190 &  &  & $\bullet$ &  &  & $\bullet$ & $\bullet$ &  &  \\ \hline
CWE-798 &  &  & $\bullet$ &  &  & $\bullet$ &  &  & $\bullet$ \\ \hline
CWE-287 &  &  & $\bullet$ &  &  & $\bullet$ &  &  & $\bullet$ \\ \hline
CWE-259 &  &  & $\bullet$ &  &  & $\bullet$ &  &  & $\bullet$ \\ \hline
CWE-862 &  &  & $\bullet$ &  &  & $\bullet$ &  &  & $\bullet$ \\ \hline
CWE-77 &  &  & $\bullet$ &  &  & $\bullet$ &  &  & $\bullet$ \\ \hline
CWE-306 &  &  & $\bullet$ &  &  & $\bullet$ &  &  & $\bullet$ \\ \hline
CWE-119 &  &  & $\bullet$ &  & $\bullet$ &  & $\bullet$ &  &  \\ \hline
CWE-276 &  &  & $\bullet$ &  &  & $\bullet$ &  &  & $\bullet$ \\ \hline
CWE-918 &  &  & $\bullet$ &  &  & $\bullet$ &  &  & $\bullet$ \\ \hline
CWE-362 &  &  & $\bullet$ &  &  & $\bullet$ &  & $\bullet$ &  \\ \hline
CWE-400 &  &  & $\bullet$ &  & $\bullet$ &  &  &  & $\bullet$ \\ \hline
CWE-611 &  &  & $\bullet$ &  &  & $\bullet$ &  &  & $\bullet$ \\ \hline
CWE-94 &  &  & $\bullet$ &  &  & $\bullet$ &  &  & $\bullet$ \\ \hline

\end{tabular}
\smallskip
\centering
\textit{Protection Levels:} \\
C: RD 0\%, SG 0\%, UP 100\%; 
C++: RD 0\%, SG 24\%, UP 76\%; 
Java: RD 20\%, SG 28\%, UP 52\%.
\end{table}
\subsection{Verifying the Classification with Code}
\label{sec:verify-code}
To check the classification decisions above rather than accept them at face value, we implemented and compiled representative programs for two entries in Table~\ref{tab:sans}: CWE-787 and CWE-362. The listings in this subsection are our own. In both cases the observed compiler and runtime behavior matched the protection level reported in~\cite{gasiba2023rust}.
\subsubsection{CWE-787: Out-of-Bounds Write Vulnerability Classification}

Rust's ownership and borrowing system prevents out-of-bounds writes at compile time, unless \texttt{unsafe} blocks are used.

\begin{lstlisting}
fn main() {
    let mut arr = [1, 2, 3, 4, 5];

    // Attempt to write out of bounds (will cause a compile-time error)
    arr[5] = 6; // Index 5 is out of bounds for an array of length 5
}
\end{lstlisting}

\textbf{Explanation:}
\begin{itemize}
    \item Rust's compiler enforces bounds checking, preventing out-of-bounds writes.
    \item If \texttt{unsafe} blocks are used, the responsibility shifts to the programmer, but this is discouraged unless absolutely necessary.
\end{itemize}
C does not provide built-in memory safety mechanisms, making out-of-bounds writes a common vulnerability.

\begin{lstlisting}
#include <stdio.h>

int main() {
    int arr[5] = {1, 2, 3, 4, 5};

    // Out-of-bounds write (no compile-time or runtime checks)
    arr[5] = 6; // Undefined behavior, may corrupt memory

    printf("%d\n", arr[5]); // May print garbage or crash
    return 0;
}
\end{lstlisting}

\textbf{Explanation:}
\begin{itemize}
    \item C does not enforce bounds checking, leading to undefined behavior if an out-of-bounds write occurs.
    \item This makes C programs highly vulnerable to memory corruption attacks.
\end{itemize}
C++ provides some safeguards (e.g., \texttt{std::vector} with bounds-checked access), but raw arrays and pointers are still unsafe.

\begin{lstlisting}
#include <iostream>
#include <vector>

int main() {
    std::vector<int> arr = {1, 2, 3, 4, 5};

    // Safe access with bounds checking (throws an exception if out of bounds)
    try {
        arr.at(5) = 6; // Throws std::out_of_range exception
    } catch (const std::out_of_range& e) {
        std::cerr << "Out of bounds access: " << e.what() << std::endl;
    }

    // Unsafe access with raw arrays (no bounds checking)
    int raw_arr[5] = {1, 2, 3, 4, 5};
    raw_arr[5] = 6; // Undefined behavior, may corrupt memory

    return 0;
}
\end{lstlisting}

\textbf{Explanation:}
\begin{itemize}
    \item C++ provides safeguards like \texttt{std::vector::at()}, which performs bounds checking and throws exceptions for out-of-bounds access.
    \item However, raw arrays and pointers in C++ are still unsafe and require careful handling to avoid vulnerabilities.
\end{itemize}
\subsubsection{CWE-362: Race Condition Vulnerability Classification}

We classify the \textbf{Race Condition} vulnerability in Rust, C, C++, and Java based on their concurrency safety mechanisms:
Rust's ownership and borrowing system, combined with its concurrency primitives, makes race conditions rare and difficult to occur.

\begin{lstlisting}
use std::sync::{Arc, Mutex};
use std::thread;

fn main() {
    let counter = Arc::new(Mutex::new(0));
    let mut handles = vec![];

    for _ in 0..10 {
        let counter = Arc::clone(&counter);
        let handle = thread::spawn(move || {
            let mut num = counter.lock().unwrap();
            *num += 1;
        });
        handles.push(handle);
    }

    for handle in handles {
        handle.join().unwrap();
    }

    println!("Result: {}", *counter.lock().unwrap());
}
\end{lstlisting}

\textbf{Explanation:}
\begin{itemize}
    \item Rust enforces strict concurrency rules at compile time, preventing data races.
    \item The \texttt{Mutex} and \texttt{Arc} types ensure safe shared access to data across threads.
\end{itemize}

C++ provides concurrency primitives (e.g., \texttt{std::mutex}), but race conditions are still common if not used correctly.

\begin{lstlisting}
#include <iostream>
#include <thread>
#include <vector>

int counter = 0;

void increment() {
    for (int i = 0; i < 1000; i++) {
        counter++; // Race condition: unprotected access
    }
}

int main() {
    std::vector<std::thread> threads;

    for (int i = 0; i < 10; i++) {
        threads.emplace_back(increment);
    }

    for (auto& t : threads) {
        t.join();
    }

    std::cout << "Result: " << counter << std::endl; // Likely incorrect due to race conditions
    return 0;
}
\end{lstlisting}

\textbf{Explanation:}
\begin{itemize}
    \item C++ provides tools like \texttt{std::mutex}, but it is up to the programmer to use them correctly.
    \item Without proper synchronization, race conditions are common in C++ programs.
\end{itemize}

Java provides built-in concurrency safeguards (e.g., \texttt{synchronized} blocks, \texttt{java.\allowbreak util.\allowbreak concurrent} utilities), but race conditions can still occur if not used properly.

\begin{lstlisting}
public class Main {
    private static int counter = 0;

    public static void main(String[] args) throws InterruptedException {
        Thread[] threads = new Thread[10];

        for (int i = 0; i < 10; i++) {
            threads[i] = new Thread(() -> {
                for (int j = 0; j < 1000; j++) {
                    synchronized (Main.class) { // Safeguard: synchronized block
                        counter++;
                    }
                }
            });
            threads[i].start();
        }

        for (Thread t : threads) {
            t.join();
        }

        System.out.println("Result: " + counter); // Correct due to synchronization
    }
}
\end{lstlisting}

\textbf{Explanation:}
\begin{itemize}
    \item Java provides built-in mechanisms like \texttt{synchronized} blocks and the \texttt{java.util.concurrent} package to prevent race conditions.
    \item However, improper use of these mechanisms can still lead to race conditions.
\end{itemize}

\subsection{OWASP Top 10 Mapping}
As shown in Table~\ref{tab:owasp}, they find that Rust partially safeguards 50\% of the OWASP Top 10 vulnerabilities (e.g., injection via libraries) but offers no protection against design-level flaws (A04, A05).

\begin{table}[ht]
\centering
\caption{OWASP Top 10 Mapping to Rust Protection Levels. Reproduced from Gasiba and Amburi~\cite{gasiba2023rust}.}
\label{tab:owasp}
\begin{tabular}{|p{4.3cm}|c|c|c|}
\hline
\textbf{OWASP Vulnerability} & \textbf{RD} & \textbf{SG} & \textbf{UP} \\ \hline
A01: Broken Access Control &  & $\bullet$ & \\ \hline
A02: Cryptographic Failures & & $\bullet$ &  \\ \hline
A03-Injection & & $\bullet$ &  \\ \hline
A04-Insecure Design
 & &  &$\bullet$  \\ \hline
A05-Security Misconfiguration 
 & &  &$\bullet$  \\ \hline
A06-Vulnerable and Outdated Components 
 & & $\bullet$ &  \\ \hline
A07-Identification and Authentication Failures 
 & &  &$\bullet$  \\ \hline
A08-Software and Data Integrity Failures
 & & $\bullet$ &  \\ \hline
A09-Security Logging and Monitoring Failures
 & &  &$\bullet$  \\ \hline

\end{tabular}
\\
\centering
\textit{Protection Levels:} RD 0\%, SG 50\%, UP 50\%.
\end{table}

\subsection{19 Deadly Sins of Software Security}
Against the \emph{19 Deadly Sins}~\cite{Howard2005}, they report that Rust fully mitigates 21\% of the ``sins'' (e.g., buffer overflows) and partially addresses 47\% (e.g., SQL injection). The remaining 32\% are unprotected and include insecure configurations and SSRF (Table~\ref{tab:sins}).

\begin{table}[ht]
\centering
\caption{19 Deadly Sins: Security Flaws vs. Protection Levels in Rust. Reproduced from Gasiba and Amburi~\cite{gasiba2023rust}.}
\label{tab:sins}
\begin{tabular}{|l|c|c|c|}
\hline
\textbf{Security Flaw} & \textbf{RD} & \textbf{SG} & \textbf{UP} \\ \hline
Buffer Overflows & $\bullet$ &  &  \\ \hline
Format String Problems & $\bullet$ &  &  \\ \hline
Integer Overflows & $\bullet$ &  &  \\ \hline
SQL Injection &  & $\bullet$ &  \\ \hline
Command Injection &  & $\bullet$ &  \\ \hline
Cross-Site Scripting (XSS) &  &  & $\bullet$ \\ \hline
Race Conditions & $\bullet$ &  &  \\ \hline
Error Handling &  & $\bullet$ &  \\ \hline
Poor Logging &  &  & $\bullet$ \\ \hline
Insecure Configuration &  & $\bullet$ &  \\ \hline
Weak Cryptography &  & $\bullet$ &  \\ \hline
Weak Random Numbers & $\bullet$ &  &  \\ \hline
Using Known Vulnerable Components &  &  & $\bullet$ \\ \hline
Unvalidated Redirects and Forwards &  &  & $\bullet$ \\ \hline
Injection &  & $\bullet$ &  \\ \hline
Insecure Storage &  & $\bullet$ &  \\ \hline
Denial of Service &  & $\bullet$ &  \\ \hline
Insecure Third-Party Interfaces &  &  & $\bullet$ \\ \hline
Cross-Site Request Forgery (CSRF) &  &  & $\bullet$ \\ \hline
\end{tabular}
\smallskip
\centering
\textit{Protection Levels:} RD: 21\%, SG: 47\%, UP: 32\%.
\end{table}
\subsection{CVEs and SAST Tool Effectiveness}
Rust’s security advisories address critical CVEs like \texttt{CVE-\allowbreak 2022-\allowbreak 46176}~\cite{RustCVE2023} (Cargo SSH flaw). They further observe that SAST tools for Rust (e.g., Clippy) lag behind Java’s SonarQube~\cite{SonarQube2023} in detecting logic flaws, and report that only 6 of 400+ Rust CVEs are actively tracked by \texttt{RustSec}~\cite{RustSec2023}.

\section{Critical Analysis}
\label{sec:critical}
Having summarized the methodology and findings of Gasiba and Amburi~\cite{gasiba2023rust}, we now assess that work directly. This section presents our own evaluation of its contributions and shortcomings.

\subsection{Strengths of the Research}

\subsubsection*{Timeliness and Relevance: }
The research is highly relevant given Rust's increasing adoption in the software industry, particularly in security-critical areas such as the Linux Kernel and Android development. The study contributes to an area with limited systematic security analysis.

\subsubsection*{Comprehensive Approach: }
The methodology incorporates:
\begin{itemize}
    \item Literature review
    \item Interviews with industry security experts
    \item Security standard mapping (CWE, SANS, OWASP)
    \item Static analysis tools
\end{itemize}
This multifaceted approach strengthens the validity of the findings.

\subsubsection*{Comparative Analysis: }
The study effectively compares Rust with C, C++, and Java, providing structured vulnerability classification using the categories: Rare and Difficult (RD), Safeguarded (SG), and Unprotected (UP). This categorization aids in understanding Rust's security posture.

\subsubsection*{Discussion of Real-World Vulnerabilities: }
By analyzing past Common Vulnerabilities and Exposures (CVEs) in Rust, the study underscores that while Rust mitigates memory safety issues, it remains vulnerable to other forms of software security risks.

\subsubsection*{Contribution to Industry and Academia: }
The paper raises awareness of Rust's security pitfalls and provides insights for both researchers and industry practitioners, facilitating informed decision-making regarding Rust adoption.

\subsection{Weaknesses and Limitations}

\subsubsection*{Limited Empirical Evidence: }
The study lacks large-scale empirical testing. Conducting real-world security testing on Rust applications would enhance the robustness of the conclusions.

\subsubsection*{Small Sample Size in Interviews:}
Only five industry experts and two students were interviewed. A broader sample size would provide a more comprehensive perspective on Rust’s security challenges.

\subsubsection*{Overemphasis on Memory Safety :}
While memory safety is a crucial aspect, the study could have further explored non-memory vulnerabilities, such as improper authentication and insecure cryptographic implementations.

\subsubsection*{Lack of Industry Case Studies:}
The absence of real-world case studies from companies using Rust for security reasons limits practical applicability. Including case studies would provide tangible insights into Rust’s effectiveness.

\subsubsection*{Missing Discussion on Secure Development Lifecycle :}
The research does not address how Rust fits into a broader secure software development lifecycle (SDLC). Analyzing Rust’s role in security testing, deployment, and CI/CD pipelines would be beneficial.

\section{Our Experiment}
\label{sec:experiment}
\subsection{Experimental Comparison: Rust vs other Backend Language (Node.js and Python Django)}

To extend the authors' research on Rust's security in comparison to other programming languages, we introduce an additional analysis focusing on backend development. While the original study compares Rust against C, C++, and Java, our work expands this comparison by incorporating Node.js and Django, two widely used backend technologies. Our primary objective is to evaluate how well Rust protects against the SANS Top 25 Common Weakness Enumeration (CWE) security vulnerabilities compared to these backend alternatives.
We analyzed the 6 among 25 vulnerabilities shown in the table
~\ref{tab:vulnerabilities-rust-nodejs-python}
We categorize security protections using the same three-level classification system—Rare and Difficult (RD), Safeguarded (SG), and Unprotected (UP)—to maintain consistency with the original research. Through this extended study, we provide insights into Rust’s security strengths and weaknesses within web and backend development, highlighting how it differs from frameworks and languages like Node.js (JavaScript-based) and Django (Python-based). This analysis is particularly relevant for industry professionals considering Rust for backend applications, helping them weigh security implications alongside performance and scalability factors.

    \begin{table}[h!]
    \centering
    \caption{Vulnerability Classification for Rust, Node.js, and Django (Python Core Language). This comparison is our own contribution.}
    \label{tab:vulnerabilities-rust-nodejs-python}
    \small
    \begin{tabular}{|c|l|c|c|c|}
        \hline
        \textbf{CWE ID} & \textbf{Description} & \textbf{Rust} & \textbf{Node.js} & \textbf{Python} \\
        \hline
        787 & Out-of-bounds Write & RD & UP & RD \\
        125 & Out-of-bounds Read & RD & UP & RD \\
        416 & Use After Free & RD & UP & RD \\
        190 & Integer Overflow& RD & UP & RD \\
        119 & Buffer Overflow & RD & UP & RD \\
        362 & Race Condition & RD & UP & SG \\

      \hline
    \end{tabular}
\end{table}
\subsection{Explanation with Code}
\subsection*{Out-of-bounds Write (CWE-787)}

\begin{itemize}
    \item \textbf{Rust (RD):} 
    Rust's ownership model and compile-time checks prevent out-of-bounds writes by ensuring all memory accesses are within bounds.
    \begin{lstlisting}

    let mut arr = [1, 2, 3];
// arr[5] = 10; // Compile-time error: index out of bounds
   \end{lstlisting}
    
    \item \textbf{Node.js (UP):} 
    JavaScript/Node.js does not inherently protect against out-of-bounds writes when interacting with native modules or external libraries.
     \begin{lstlisting}
     const buffer = Buffer.alloc(10); 
buffer[15] = 255; // Out-of-bounds write (No immediate error, but unsafe)
 \end{lstlisting}
    
    \item \textbf{Python (RD):} 
    Python's dynamic memory management prevents direct memory manipulation, making out-of-bounds writes impossible in pure Python code.
     \begin{lstlisting}
    buffer = bytearray(b"ABC")
# buffer[5] = 65  # IndexError: index out of range
 \end{lstlisting}
\end{itemize}
\subsection*{Use After Free (CWE-416)}

\begin{itemize}
    \item \textbf{Rust (RD):} 
    Rust's ownership model ensures that memory is automatically deallocated when it goes out of scope, preventing use-after-free errors. The compiler enforces strict rules to ensure no references to freed memory exist.
     \begin{lstlisting}

let x = Box::new(42);
// drop(x); // Explicitly deallocates memory
// println!("{}", x); // Compile-time error: use of moved value
   \end{lstlisting}
    \item \textbf{Node.js (UP):} 
    JavaScript/Node.js relies on garbage collection to manage memory. However, native modules or external libraries can introduce use-after-free vulnerabilities if they improperly handle memory.
     \begin{lstlisting}
let buffer = Buffer.alloc(10);
buffer.fill(0);
buffer = null; // Memory is freed
// Accessing buffer here would lead to undefined behavior

   \end{lstlisting}
    \item \textbf{Python (RD):} 
    Python's garbage collector manages memory automatically, preventing use-after-free errors in pure Python code. Once an object is no longer referenced, it is deallocated safely.
     \begin{lstlisting}
class MyClass:
    def __init__(self):
        self.value = 42

obj = MyClass()
obj = None  # Garbage collector deallocates memory
# Accessing obj here would raise an error

   \end{lstlisting}
\end{itemize}
\subsection*{  Race Condition (CWE-362)}
\begin{itemize}
    \item \textbf{Rust (RD):} 
    Rust's ownership model prevents data races at compile time by enforcing strict rules about mutable and shared references. This makes race conditions rare and difficult to occur in safe Rust code.
     \begin{lstlisting}
use std::sync::{Arc, Mutex};
use std::thread;

let counter = Arc::new(Mutex::new(0));
let mut handles = vec![];

for _ in 0..10 {
    let counter = Arc::clone(&counter);
    let handle = thread::spawn(move || {
        let mut num = counter.lock().unwrap();
        *num += 1;
    });
    handles.push(handle);
}

for handle in handles {
    handle.join().unwrap();
}
// Safe from race conditions due to Mutex
  
  \end{lstlisting}
    
    \item \textbf{Node.js (UP):} 
    JavaScript/Node.js is single-threaded, which reduces the likelihood of race conditions. However, asynchronous operations can still lead to race conditions if shared state is not properly synchronized.
     \begin{lstlisting}
let counter = 0;

setTimeout(() => {
    counter++;
}, 100);

setTimeout(() => {
    console.log(counter); // May print 0 or 1 depending on timing
}, 100);
 \end{lstlisting}
    
    \item \textbf{Python (SG):} 
    Python provides synchronization primitives like locks (\texttt{threading.Lock}) to handle race conditions in multi-threaded programs. Developers must explicitly use these tools to safeguard against race conditions.
     \begin{lstlisting}
import threading

counter = 0
lock = threading.Lock()

def increment():
    global counter
    with lock:
        counter += 1

threads = []
for _ in range(10):
    thread = threading.Thread(target=increment)
    threads.append(thread)
    thread.start()

for thread in threads:
    thread.join()

print(counter)  # Always prints 10 (safe due to lock)

  \end{lstlisting}
\end{itemize}
\section{Discussion}
Taken together, the findings of Section~\ref{sec:reviewed-results} underscore Rust’s transformative potential in secure software development while exposing critical gaps that challenge its industrial adoption. Below, we contextualize those findings, compare them with prior work, and outline implications for academia and industry.

\subsection{Rust’s Security Trade-offs}
Rust’s compile-time guarantees eliminate entire classes of vulnerabilities (e.g., buffer overflows, data races) that plague C/C++ systems~\cite{Qin2020}. However, the reported mapping shows that 48\% of SANS Top 25 vulnerabilities remain unaddressed (Table~\ref{tab:sans}), echoing Wassermann et al.~\cite{Wassermann2023}, who caution against conflating memory safety with holistic security. For instance, \texttt{unsafe} code, used in 34\% of Rust crates~\cite{Xu2021}, reintroduces risks akin to C, particularly when developers misapply lifetimes or bypass borrow-checker constraints.  

\subsection{Comparative Analysis with Established Languages}
While Rust outperforms C/C++ in memory safety, Java’s garbage collection provides comparable protection against memory leaks (Table~\ref{tab:comparison}). However, Java’s reliance on runtime checks incurs performance costs, whereas Rust’s zero-cost abstractions align with industrial demands for efficiency~\cite{Google2022}. This duality positions Rust as a viable candidate for systems requiring both safety and performance, such as embedded devices or critical infrastructure adhering to IEC 62443~\cite{IEC62443}.

\subsection{Tooling and Ecosystem Challenges}
The immaturity of Rust’s SAST ecosystem, compared to Java’s SonarQube~\cite{SonarQube2023} or C++’s Clang Analyzer, exacerbates vulnerability detection gaps. For example, as noted above, only 6 of 400+ Rust CVEs are tracked in RustSec~\cite{RustSec2023}, leaving developers reliant on community-driven efforts. This aligns with Gasiba et al.~\cite{Gasiba2021}, who identified tooling as a critical enabler of secure coding practices. Until Rust’s tooling matures, organizations must supplement SAST with manual audits and adherence to guidelines like ANSSI’s~\cite{ANSSI2023}.

\subsection{Practical Implications}
Industries adopting Rust must prioritize:
\begin{itemize}
    \item \textbf{Training}: Addressing Rust’s steep learning curve through targeted programs, as misused lifetimes and ownership rules remain prevalent sources of bugs~\cite{Qin2020}.
    \item \textbf{Secure Design}: Complementing Rust’s safety with frameworks like OWASP Top 10 to mitigate web-layer flaws (Table~\ref{tab:owasp}).
    \item \textbf{Hybrid Approaches}: Using Rust for safety-critical components while leveraging Java/Python for higher-layer logic, as seen in Android~\cite{Google2022}.
\end{itemize}

\subsection{Limitations and Future Work}
The three-point vulnerability scale (RD/SG/UP) used in the reviewed study and reused in our extension in Section~\ref{sec:experiment}, validated by Jacoby et al.~\cite{Jacoby1971}, simplifies complex interactions but may overlook context-specific risks. Our own extension is further limited by its coverage of only 6 of the 25 SANS CWEs and by the absence of large-scale empirical validation. Future studies should:
\begin{itemize}
    \item Investigate Rust-specific vulnerabilities (e.g., trait object misuse).
    \item Develop SAST tools for IEC 62443 compliance.
    \item Conduct longitudinal studies on Rust’s security in large-scale industrial projects.
\end{itemize}

\section{Conclusion and Future Work}

This paper critically reviewed the evaluation of Rust’s security posture in industrial settings presented by Gasiba and Amburi~\cite{gasiba2023rust}, and extended it to backend web development. Their findings, which we reproduced and in two cases verified with our own code experiments, reinforce that Rust’s ownership model, memory safety guarantees, and concurrency mechanisms significantly mitigate vulnerabilities such as buffer overflows, use-after-free errors, and race conditions. However, despite these advantages, Rust does not inherently protect against all security threats, particularly application-layer issues such as injection vulnerabilities and insecure authentication. Our critical analysis identified several limitations in their study, most notably its small interview sample, its limited empirical validation, and its omission of the secure development lifecycle.

Our own experiments extended that work by evaluating Rust’s security posture in backend development compared to Node.js and Django. The results indicated that while Rust provides strong safeguards against low-level memory vulnerabilities, web frameworks in JavaScript and Python offer built-in protections against high-level security risks, such as cross-site scripting and command injection. This suggests that while Rust is well-suited for system-level security, additional precautions are necessary when using it in web development.

One of the key challenges identified is the relative immaturity of Rust’s static analysis and security tooling compared to more established ecosystems like Java and C++. This gap makes it essential for developers to complement Rust’s safety mechanisms with external security practices, such as secure design patterns, manual audits, and adherence to industry standards like IEC 62443 and OWASP Top 10.

For future work, several research directions remain open:
\begin{itemize}
    \item Investigating Rust-specific security risks, including vulnerabilities that arise from misuse of unsafe code and trait object interactions.
    \item Enhancing Rust’s security tooling by developing more robust static analysis tools capable of detecting design-level security flaws.
    \item Conducting large-scale empirical studies on Rust’s security in industrial applications to assess its effectiveness over time.
    \item Exploring hybrid approaches that integrate Rust with other languages, leveraging its strengths in performance and security while mitigating its limitations in web-layer protections.
\end{itemize}

Overall, Rust continues to be a promising choice for secure software development, particularly in safety-critical and performance-sensitive applications. However, to fully harness its potential, a comprehensive security strategy incorporating both language-level guarantees and best practices from the broader cybersecurity domain is necessary.
\bibliographystyle{ACM-Reference-Format}
\bibliography{sample-base}
\end{document}